\documentclass[12pt]{article}
\pdfoutput=1
\usepackage{epsf,amsfonts,amssymb,epsfig,amsmath,mathtools,graphics,slashed}
\usepackage{hep,graphicx,subfig} 
\usepackage{rotating}
\usepackage{jheppub}

\graphicspath{./figs/}

\newcommand{\nn}{\notag \\}

\makeatletter

\@addtoreset{equation}{section}
\makeatother

\author[1]{Aron Jansen}
\emailAdd{a.p.jansen@icc.ub.edu}
\affiliation[1]{Departament de Fisica Quantica i Astrofisica and Institut de Ciencies del Cosmos, \\
Universitat de Barcelona, Marti i Franques 1, ES-08028, Barcelona, Spain.}
\author[2]{Christiana Pantelidou}
\emailAdd{cpantelidou@maths.tcd.ie}
\affiliation[2]{School of Mathematics, Trinity College Dublin, Dublin 2, Ireland.}

\title{Quasinormal modes in charged fluids at complex momentum}

\abstract{
We investigate the convergence of relativistic hydrodynamics in charged fluids, within the framework of holography. On the one hand, we consider the analyticity properties of the dispersion relations of the hydrodynamic modes on the complex frequency and momentum plane and on the other hand, we perform a perturbative expansion of the dispersion relations in small momenta to a very high order. We see that the locations of the branch points extracted using the first approach are in good quantitative agreement with the radius of convergence extracted perturbatively. We see that for different values of the charge, different types of pole collisions set the radius of convergence. 
In all hydrodynamic channels the latter turns out to be finite, while it  goes to zero at extremality for all hydrodynamic modes.
Furthermore, we also establish the phenomenon of pole-skipping for the Reissner-Nordstr\"{o}m black hole, and we find that the value of the momentum for which this phenomenon occurs need not be within the radius of convergence of hydrodynamics.
}

\notoc

\begin{document}
\maketitle

\setcounter{equation}{0}

\section{Introduction}

Charged hydrodynamics is an effective theory of fluids describing the evolution of conserved quantities, namely energy, momentum and local charge density, when near equilibrium. In the relativistic setting, which is the focus here, one starts with the perfect fluid stress tensor and the electric current  and systematically corrects them by adding higher gradient terms. Each of the latter enters multiplied by a transport coefficient, which is simply a function of the equilibrium temperature and chemical potential. Fixing these transport coefficients is equivalent to fixing a microscopic theory. This is a very universal statement, which makes hydrodynamics a very powerful approach.

A charged homogeneous and isotropic relativistic fluid which exhibits a hydrodynamic limit supports collective excitations in the form of shear, sound and charge diffusion modes. These modes arise by considering linearised fluctuations of the local hydrodynamic variables around the equilibrium configuration. 
They describe the transverse gauge and velocity fluctuations, the propagation of the longitudinal velocity fluctuation and the fluctuation of the energy and the pressure, and the diffusion of the fluctuations of the charge density.
The frequency of these modes is controlled by their momentum, $k$, through the dispersion relations 
\begin{align}\label{eq:dispersions}
&\omega_{shear}(k)=- i\, D\, k^2+\dots\,,\nn
&\omega_{sound}(k)=\pm v_s\, k- i \frac{\Gamma_s}{2} k^2+\dots\,,\nn
&\omega_{charg}(k)=- i \, D_R\, k^2+\dots\,,
\end{align}
where $v_s$ is the speed of sound, $D=\eta/(\epsilon+p)$, $D_R$ is the charge diffusion constants and $\Gamma_s=[\zeta+\tfrac{2d-4}{d-1}\eta]/(\epsilon+p)$. Here $d$ is the number of spacetime dimensions, $\epsilon$ is the energy density, $p$ is the pressure, $\eta$ is the shear viscosity and $\zeta$ is the bulk viscosity. The terms shown above come from a derivative expansion to first order, while higher order terms in the derivative expansion will appear with higher powers of $k$. For a review of relativistic hydrodynamics, we refer the  interested reader to \cite{Kovtun:2012rj}. 

In order to determine the precise values of the coefficient appearing in \eqref{eq:dispersions}, one needs to fix an underlying microscopic theory. The holographic framework provides such microscopically complete theories. In particular, in this work we will focus on a holographic model given by Einstein-Maxwell theory in $AdS_{d+1}$ with the equilibrium configuration described by the Reissner-Nordstr\"{o}m black brane. The linear modes \eqref{eq:dispersions} holographically correspond to quasinormal modes (QNMs) associated with poles of retarded correlators of the conserved currents.

The aim of this paper is to further investigate the analytic properties of correlators in the complex momentum plane for a charge fluid within the framework of holography. This is an interesting direction to pursue because it will shed light on the convergence properties of hydrodynamics, its radius of convergence and the possibility of resumming it in order to extract information about the microscopic theory beyond its hydrodynamic limit.

The convergence of hydrodynamics, including determining its radius of convergence and resumming, has been addressed before in the literature. Focusing on boost-invariant flows, where the additional symmetries of the problem lead to major simplifications, \cite{Heller:2013fn} showed that the late proper time  expansion is typically asymptotic. The issue with convergence comes because of contributions from short-lived non-hydrodynamic modes in the spectrum. This was concluded through Borel-Pade and resurgence techniques which have been used to resum the perturbative series in order to extract information about these non-hydrodynamic modes \cite{Heller:2015dha,Basar:2015ava,Aniceto:2015mto,Florkowski:2016zsi,Denicol:2016bjh,Heller:2016rtz,Florkowski:2017olj,Spalinski:2017mel,Casalderrey-Solana:2017zyh,Heller:2018qvh}. Resummation of hydrodynamics was also discussed in the context of the kinetic theory \cite{Kurkela:2017xis} and cosmological models \cite{Buchel:2016cbj}, while in the context of the fluid/gravity correspondence it was investigated in \cite{Bu:2014ena}, building on \cite{Bu:2014sia}.

Focusing on linearized hydrodynamics but not assuming any additional symmetries, the convergence of the hydrodynamic series in the shear-diffusion channel in $d=3$ was investigated in a holographic model involving a dual $AdS_4$-Reissner-Nordstr\"{o}m black brane \cite{Withers:2018srf} for a particular value of the charge $Q$. Using the coefficients of a small momentum  expansion of the dispersion relation, \cite{Withers:2018srf} concluded that the radius of convergence of the shear-mode series was finite and given by
\begin{equation}
k^{shear}_c=\frac{\epsilon+p}{2 \mu \sqrt{\eta}}\,.
\end{equation}
The obstruction to convergence was associated to a (square-root) branch point at purely imaginary momentum, $k=i\, k^{shear}_c$, caused by a collision with a gapped QNM. This non-analyticity was the closest to $k=0$ and thus was the one to set the radius of convergence. \cite{Withers:2018srf} proceeded then to Pade-resum the asymptotic series and extend it beyond the branch point onto the first non-hydrodynamic sheet. Going beyond and onto the second non-hydrodynamic sheet required another resummation due to the presence of a second branch cut and eventually led to the recovery of the first non-hydrodynamic mode. Note that the formula above is inversely proportional to the chemical potential, $\mu$, and thus if one extrapolates to $\mu=0$,  it suggests that this type of non-analyticity takes place at infinity. Of course this does not exclude the possibility of other critical points appearing at smaller value of the momenta, which as we will see is the case here.

In the absence of a chemical potential, \cite{Grozdanov:2019uhi,Grozdanov:2019kge,Natsuume:2019xcy} investigated the convergence of the hydrodynamic expansion in several cases, including a holographic model with $d=4$ boundary dimensions involving the Schwarzschild-$AdS_5$ black brane. Once again, the radius of convergence was determined, but this time through a computation involving complex hydrodynamic spectral curves. The spectral curve is defined as the curve given by 
\begin{align}
F(k^2,\omega)=0\,,
\end{align}
where $F$ is proportional to the inverse of the corresponding retarded correlator, and its critical points are given by
\begin{align}
F(k_c^2,\omega_c)=0\,,\qquad \frac{\partial F}{\partial \omega}(k_c^2,\omega_c)=0\,.
\end{align}
In order to determine the radius of convergence of hydrodynamics one should focus on the hydrodynamic spectral curve, which are spectral curves that contain a solution with the property $\omega(k\to 0)\to 0$, giving rise to the dispersion relation of a hydrodynamic mode. Using this approach, \cite{Grozdanov:2019uhi,Grozdanov:2019kge} concluded that the radius of convergence of hydrodynamics in their system was finite.

A natural question to ask is how the two approaches for determining the radius of convergence of hydrodynamics compare and what is the complete picture for the radius of convergence as a function of the charge. The results that currently exist in the literature using the two approaches correspond to different values of the charge, $Q$, and different dimensions, $d$, and thus direct comparison is not possible. 
However naive extrapolation of the result of \cite{Withers:2018srf} gives an apparent disagreement, and this has led to confusion in the literature, which we clear up here.
In this paper, we carry out a thorough study of the radius of convergence as a function of the charge for both $d=3$ and $d=4$ using both methods. As we will see in section \ref{sec:NonP} and \ref{sec:Pert}, the structure of the QNMs of the Reissner-Nordstr\"{o}m black hole is particularly complex, involving different types of pole collisions at different values of the charge. The consequence of this is that the analytic branch of pole collisions studied in \cite{Withers:2018srf} is dominant only for intermediate values of the charge, while a different collision sets the radius of convergence at neutrality, rendering it finite. 

More recently, \cite{Heller:2020uuy} viewed the asymptotic nature of the real space derivative expansion in relativistic hydrodynamics in terms of the state about which this expansion is performed, rather than a generic property of the hydrodynamic gradient expansion itself. In particular, they showed that the radius of convergence of the gradient expansion is finite only if the initial data have support only for momenta below a critical value, set by the smallest value of momentum where a hydrodynamic and non-hydrodynamic mode collide. The method developed in  \cite{Heller:2013fn} is quite general and applies across models and it's a milestone in understanding the behaviour of a large set of physical systems arising in heavy ion collisions and cosmology showing that the expansion in real space gradients is divergent.

In addition to the radius of convergence, we will explore pole-skipping for charged black branes. This phenomenon was first seen in \cite{Grozdanov:2017ajz}, where it was shown that when the hydrodynamic sound mode is driven to instability by a choice of a specific value of imaginary momentum $q=i q_*$, then the retarded two-point function exhibits an exponential growth related to chaos; the frequency and momentum are given by the holographic Lyapunov exponent and the butterfly velocity. At this special point, the residue of the retarded two-point function of the energy correlator vanishes and thus, this is a collision of a ``zero'' and a ``pole'' of the associated correlator. As such, at the pole skipping point the correlator is infinitely multivalued, depending on the way it is approached. It was later realised that pole skipping can be directly associated with absence of a unique ingoing solution close to the black brane horizon \cite{Blake:2018leo}. To-date this phenomenon is known to be more generic, in the sense that it manifests itself in all hydrodynamic and non-hydrodynamic channels  \cite{Grozdanov:2019uhi,Blake:2019otz}, and depending on whether it appears in the upper or lower frequency plane it might or might not be related to a chaotic behaviour. Holographically, it has been studied  mainly for the Schwarzschild black hole \cite{Grozdanov:2017ajz,Blake:2019otz} as well as for a neutral black hole that explicitly breaks translation invariance \cite{Blake:2018leo}. The fermionic case was considered in \cite{Ceplak:2019ymw}. From the field theory perspective, this phenomenon was investigated in \cite{Blake:2017ris} for effective field theories and in \cite{Haehl:2018izb} for conformal field theories with large central charge. Furthermore \cite{Das:2019tga} studied pole skipping for 2 dimensional CFT correlators of conserved currents and also a correlator in a BCFT. As we will see in section \ref{sec:poleSkip}, pole-skipping persists in holographic theories in the presence of a chemical potential in all channels.

A natural question to ask is if there is any relation between $q_*$ and $q_c$. This was investigated in \cite{Grozdanov:2019uhi} for the Schwarzschild BH, where $q_*<q_c$ in all channels, as well as for black holes that explicitly break translation invariance where $q_*>q_c$. This led \cite{Grozdanov:2019uhi}  to conclude that the two scales are in general unrelated, and in fact, the results we obtain here support this idea.

The remainder of this paper is structured as follows. In section \ref{sec:setup} we present the holographic model of interest, which is simply described by Einstein-Maxwell theory with a negative cosmological constant, and we briefly discuss the computation of QNMs using the formalism of master equations. In section \ref{sec:NonP} we proceed to determine the radius of convergence for the shear, sound and charge diffusion hydrodynamic modes by studying the dispersion relation of the corresponding modes in the complex frequency and momentum plane, i.e. by using spectral curves.  In section \ref{sec:Pert} we re-compute the radius of convergence through a small-$k$ expansion of the dispersion relation up to a very high order. We find that the two methods are in good quantitative agreement. Finally we establish the phenomenon of pole-skipping for the Reissner-Nordstr\"{o}m black brane in section \ref{sec:poleSkip}. We conclude with a discussion in \ref{sec:concl}. In appendix \ref{app:MasterEquations}, we give some more technical details about the decoupled master equations that we used in this computation.\\

\textbf{Note added:} During the final stages of this calculation, we became aware of the results of \cite{Abbasi:2020ykq}, with which there is some overlap. In \cite{Abbasi:2020ykq}, the authors study the radius of convergence using the approach of the spectral curves for charged fluids in $d=4$ for a certain range of charge (in our units $\tilde Q\in[0,0.57])$. This is to be contrasted with the current paper that considers both $d=3$ and $d=4$ using two different methods and for the complete range of $\tilde Q$, going from zero all the way to extremality. Given their limited range for $\tilde Q$, they have missed some structure that appears only at high values of $\tilde Q$. 
Furthermore for the range of $\tilde Q$ that was considered in \cite{Abbasi:2020ykq},  in the shear channel the authors identified only one dominant branch of pole collisions. This is in disagreement with our results, where we find that the analytic branch of \cite{Withers:2018srf} already becomes the dominant collision in the shear channel for large enough $\tilde Q$. We agree with their results for the diffusion channel. In the sound channel we only disagree with the identity of the colliding modes.
For  $0.27 < \tilde{Q} < 0.45$ we do not see this as a collision of the hydrodynamic sound and hydrodynamic diffusion modes, as the latter is already outside of its radius of convergence. For $\tilde{Q} > 0.45$ the collision of the sound mode is with a non-hydrodynamic mode in the sound channel that is diffusive (and not a gapped charge diffusion mode as they claim). 
In \cite{Abbasi:2020ykq} the authors also investigate the phenomenon of pole skipping for charged black holes, but they are interested only in the upper-half plane for sound $d=4$. Here we do a complete analysis of the phenomenon, including the Matsubara frequencies in the lower half plane, in both $d=3,4$ in all channels.

\section{Set up}\label{sec:setup}

We will consider Einstein-Maxwell theory in $D = d+1$ spacetime dimensions with action given by 
\begin{align}\label{eq:LagraEM}
S= \int&dx^{d+1}\left( R-2 \Lambda -\frac{1}{4} F_{\mu\nu}\, F^{\mu\nu}\right)
\end{align}
where $F\equiv dA$, $\Lambda=-d(d-1)/2$ is the cosmological constant and we have set the units by fixing $16 \pi G=1$. The variation of the action \eqref{eq:LagraEM} gives rise to the following field equations of motion
\begin{align}\label{eq:eom}
R_{\mu\nu}-\frac{R}{2} g_{\mu\nu}-\frac{d(d-1)}{2}g_{\mu\nu}+\frac{1}{4}\left(\frac{1}{2}\,g_{\mu\nu}F_{\lambda\rho}F^{\lambda\rho}-2F_{\mu\rho}F_{\nu}{}^{\rho} \right)&=0\,,\nn
\nabla_{\mu}F^{\mu\nu}&=0\,.
\end{align}
The above equations admit a unit-radius $AdS_{d+1}$ vacuum solution with $A=0$, which is dual to a $d$ dimensional CFT with an abelian global symmetry. Here we choose to place our boundary theory at a finite temperature, $T$, and deform it by a chemical potential, $\mu$. The corresponding backreacted solution in the bulk is then given by the AdS-Reissner-Nordstr\"{o}m (AdS-RN) black brane
\begin{align}\label{eq:norm_ansatz}
ds^{2}&=\frac{1}{u^2}\left(- f(u) dt^2+2 dt\, du +dx^2+dy^2\right),\notag\\
A&=\mu(1-u^{d-2})\, dt\,,
\end{align}
where
\begin{align}
f(u)=1-\left(1+Q^2\right)\,u^d+Q^2\,u^{2(d-1)}\,,\quad Q^2=\frac{d-2}{2(d-1)}\mu^2\,.
\end{align}
The black hole horizon is located at $u=1$ in these coordinates and the (dimensionless) temperature of the black hole is given by 
\begin{align}\label{tempexp}
\frac{T}{\mu} = \frac{1}{4\pi\,\mu}\left[d-Q^2(d-2)\right]\equiv\frac{d}{4\pi\,\mu}\,(1-\tilde Q^2)\,.
\end{align}
When $\tilde Q=0$ the system is neutral and the bulk solution \eqref{eq:norm_ansatz} reduces to the AdS-Schwarzschild black brane, while $\tilde Q=1$ corresponds to the extremal limit.  Note that \eqref{eq:norm_ansatz} is written in Eddington-Finkelstein coordinates for later convenience.


\subsection{Fluctuations around equilibrium}
We now move on to study fluctuations around the background configuration \eqref{eq:norm_ansatz}. Within the holographic framework, the modes that drive the microscopic theory to equilibrium correspond to quasinormal modes (QNMs). 
A detailed account of quasinormal spectra for the AdS-Schwarzschild and Reissner-Nordstr\"{o}m backgrounds with translationally invariant horizons was discussed previously in the literature~\cite{Kovtun:2005ev}. In particular, for the charged case it is well know that the electromagnetic and gravitational perturbations split into 3 sectors, depending on their transformation properties: the tensor (for $d>3$), the vector and the scalar sector. The scalar sector corresponds to coupled equations containing the sound and charge diffusion fluctuations, the vector are coupled equations containing  the shear and transverse gauge field fluctuations and the tensor modes are decoupled scalar equations. Out of these 5 types of fluctuations only the shear, sound and diffusion contain hydrodynamic modes, meaning modes that obey dispersion relations such that the frequency approaches zero as the momentum is decreased. These hydrodynamic modes correspond precisely to the shear, sound and diffusion modes in the hydrodynamic limit of the dual CFT at finite $T$ and $\mu$ that we are interested in.

Let us begin the computation of the QNMs by discussing in a bit more detail the equations governing the gravitational and electromagnetic perturbations corresponding to the above modes and the boundary conditions that we need to impose. We use the formalism of master equations, which was first derived for AdS-RN in~\cite{Kodama:2003kk}.  More recently in~\cite{Jansen:2019wag} this was rederived and generalized, resulting in a form that is better suited for our purposes (in particular getting the potential in the form of \ref{eq:U0}).

Very briefly the idea is to first find all the gauge invariant combinations of the perturbations, and then combine those further into what are called master scalars that satisfy Klein-Gordon equations with certain potentials.
The tensor sector contains a single master scalar $\Phi_2$, the vector sector contains two, whose equations can be decoupled by taking certain linear combinations that we will denote $\Phi_{1 \pm}$ and the same holds for the scalar sector, which we denote $\Phi_{0 \pm}$.
Each of these master scalars satisfies a master equation of the form
\begin{align}\label{eq:mastereqs}
\square \Phi_i + W_i(u) \Phi_i &= 0 \, ,
\end{align}
where $\square$ is the Laplacian on the spacetime \eqref{eq:norm_ansatz}. Without loss of generality, we decompose the master scalars into plane waves as
\begin{equation}
\Phi_i(t, u, x) = e^{- i (\omega t - k x)}  \Phi_i(u) \, ,
\end{equation}
where $\omega$ and $k$ are the frequency and momentum of the associated QNM. For convenience we work with the dimensionless quantities $\lambda = \omega / (2 \pi T)$ and $z = (k / (2 \pi T))^2$. 

We will refer to each of the decoupled equations as a channel within the sector.
So the tensor sector only has a single equation, the tensor channel.
The vector sector has two decoupled equations, which are the shear channel, corresponding to the negative sign, and the transverse channel corresponding to the positive sign.
The scalar sector contains the sound channel, corresponding to the positive sign, and the diffusion channel, corresponding to the negative sign.

The potentials can be written as
\begin{align}
W_2(u) &= 0 \\
W_{1\pm}(u) &= U_1(u) \pm \Delta_1 \frac{d}{2} u^{d}  \\
W_{0\pm}(u) &= \mathcal{D}(u)^{-2}\left( U_0(u) \pm \Delta_0 V_0(u) \right) \,  \label{eq:U0},
\end{align}
where $U_1, U_0, V_0$ and $\mathcal{D}$ are polynomials in the radial coordinate $u$, with polynomial dependence on the squared momentum $z$ and the squared charge $\tilde{Q}^2$.
We further rescale the master scalars as 
\begin{align}
\Phi_i(u) = u^{(d-1)/2} \Psi_i(u)
\end{align}
 to bring the equations into Schr\"{o}dinger-like form.
For the full details on these equations and their derivation we refer the interested reader to~\cite{Jansen:2019wag}.

Most importantly for our purposes as we shall see below,
\begin{align}
\Delta_1 &= \sqrt{\left( d - 2 + d \,\tilde{Q}^2 \right)^2 + 2 d(d-1)       z \tilde{Q}^2 (1 - \tilde{Q}^2)^2 }\, , \\
\Delta_0 &= \sqrt{\left( d - 2 + d\, \tilde{Q}^2 \right)^2 + 4 d\left(d-2\right) z \tilde{Q}^2 (1 - \tilde{Q}^2)^2 }\, .
\end{align}
Note that in the case $d= 3$ the two $\Delta$'s are equal.

Let us now move on to discuss the boundary condition that we impose. In the IR we need to impose ingoing boundary conditions, which since we are using Eddington-Finkelstein coordinates simply means regularity 
 \begin{equation}
\Psi_i= \psi_i^{(0)} + \psi_i^{(1)} (u-1)+\dots\,.
 \end{equation}  
 In the UV we require the absence of sources for the gravitational and electromagnetic perturbations. 
In terms of the master scalars this turns out to lead to mixed boundary conditions in the scalar channel when $d = 3$. In appendix \ref{app:MasterEquations} we show precisely why this is the case.


For the numerical computation of QNMs we used the Mathematica package of \cite{Jansen:2017oag}.
  

\section{Non-perturbative computation of radius of convergence}\label{sec:NonP} 

The aim of this section is to extract the radius of convergence of the hydrodynamic expansion for the shear, sound and diffusion collective excitations by studying the spectral curves of the corresponding QNMs and identifying points of non-analyticity. We carry out this computation for both $d=3$ and $d=4$, and we scan the whole range $\tilde Q\in[0,1]$, going from the chargeless case all the way to extremality. We consider slices of the hydrodynamic spectral curve in the complex frequency, $\lambda$, and momentum, $z$, plane along which $|z|=const$. In this context non-analyticities correspond to sudden changes in the shape of these curves caused by collisions of the hydrodynamic modes with gapped ones.

Currently there is no formal proof relating the radius of convergence of the hydrodynamic expansion with the non-analyticities on the complex momentum plane described above. However empirically, we do expect the branch points to set the radius of convergence \cite{Withers:2018srf}. Some arguments lending support in this directions are discussed in the supplementary material of \cite{Heller:2020uuy}. As we will see here this empirical expectations holds in all the cases that we checked. As such we will slightly abuse terminology to refer to the location of the branch points as the radius of convergence extracted from spectral curve considerations.

\begin{figure}[h!]
\centering
\includegraphics[width=1.\linewidth]{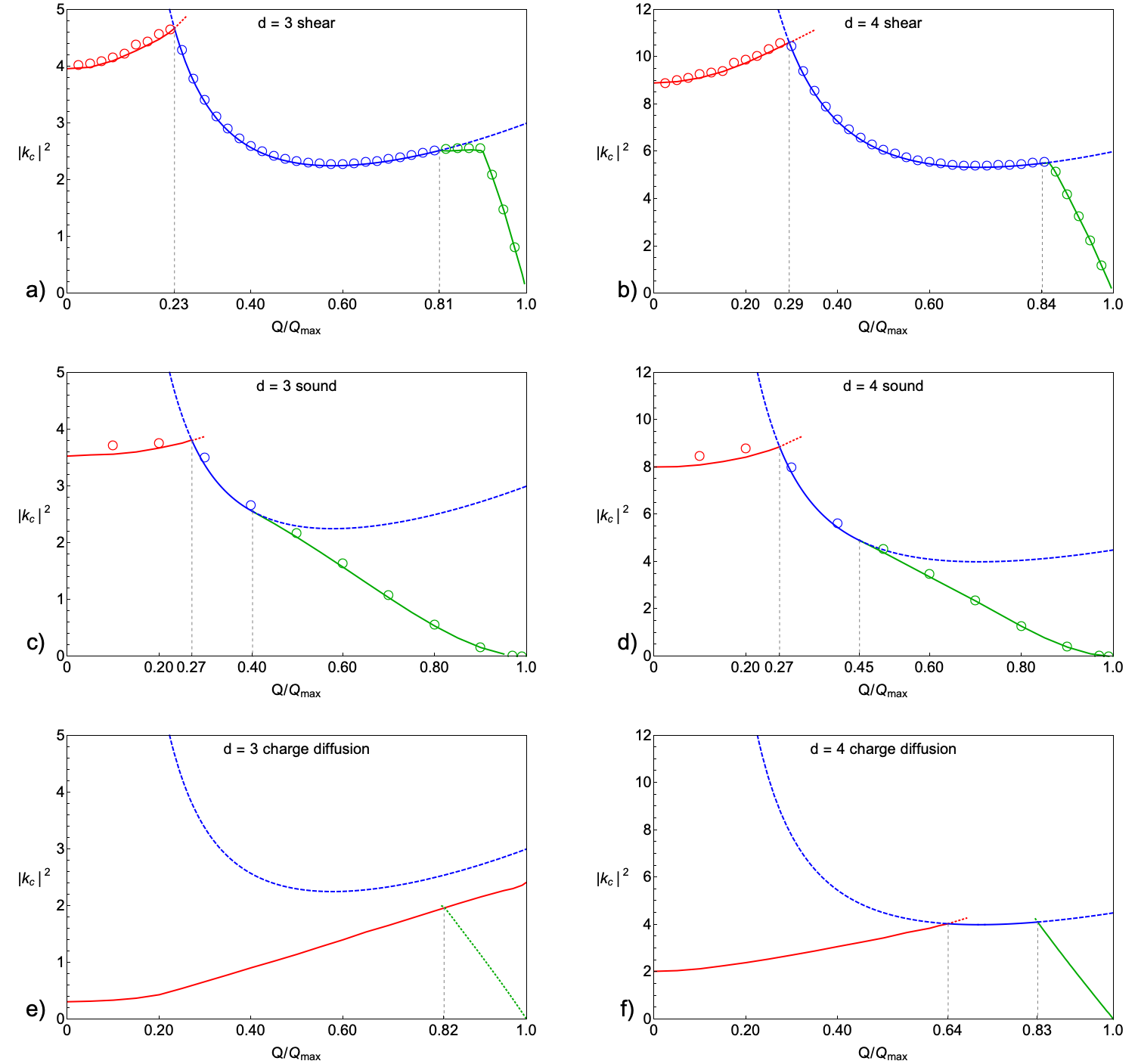}
\caption{Radius of convergence as a function of the charge in the system for  the hydrodynamic modes in our system, namely shear, sound and charge diffusion in $d=3$ and $d=4$. Each line on these plots corresponds to a different branch of pole collisions, with the solid (dashed) lines indicating where each branch is dominant (subdominant). As $\tilde Q=Q/Q_{max}$ is varied, different branches are responsible for setting the radius of convergence. Transition points are indicated with dashed vertical lines. The circles correspond to the results for the radius of convergence obtained using a perturbative expansion of the corresponding dispersion relations in small momenta up to a very high order, as discussed in section \ref{sec:Pert}. We see that the two methods are in good quantitative agreement.}
\label{fig:RC2}
\end{figure}

A careful analysis of the spectral curves of the three hydrodynamic modes contained in our system led to the results  shown in figure  \ref{fig:RC2}.  In the panels (a)-(b),(c)-(d) and (e)-(f) we show the values of the momenta at which the spectral curves of the hydrodynamic shear, sound and diffusion modes become non-analytic in $d=3$ and $d=4$ dimensions, respectively. In each case there are several branches that appear over different ranges of the charge $\tilde Q$. 
The radius of convergence is then set by the smallest absolute value of momentum at which a pole collision occurs, depicted in the plots with solid lines. The dashed lines correspond to the same collisions but in regions where they are no longer dominant.
The circles in the plots for the sound and shear correspond to the results  for the radius of convergence obtained through a perturbative calculation in small-$k$ and we will come back to them in section \ref{sec:Pert}.

In particular, we see that the radius of convergence for all the hydrodynamic modes is finite at $\tilde Q=0$ as expected from the analysis of  \cite{Grozdanov:2019uhi,Grozdanov:2019kge}, and in quantitative agreement those results in the $d=4$ case. Interestingly, in the extremal limit, $\tilde Q \to 1$, the radius of convergence for all hydrodynamic modes goes to zero. 
This matches nicely the results of \cite{Edalati:2010hk,Edalati:2010pn}, where they consider shear and sound QNMs for the extremal RN black brane in $AdS_4$ and they found a branch cut along the negative imaginary frequency axis.

\begin{figure}[h!]
\centering
\includegraphics[width=1\linewidth]{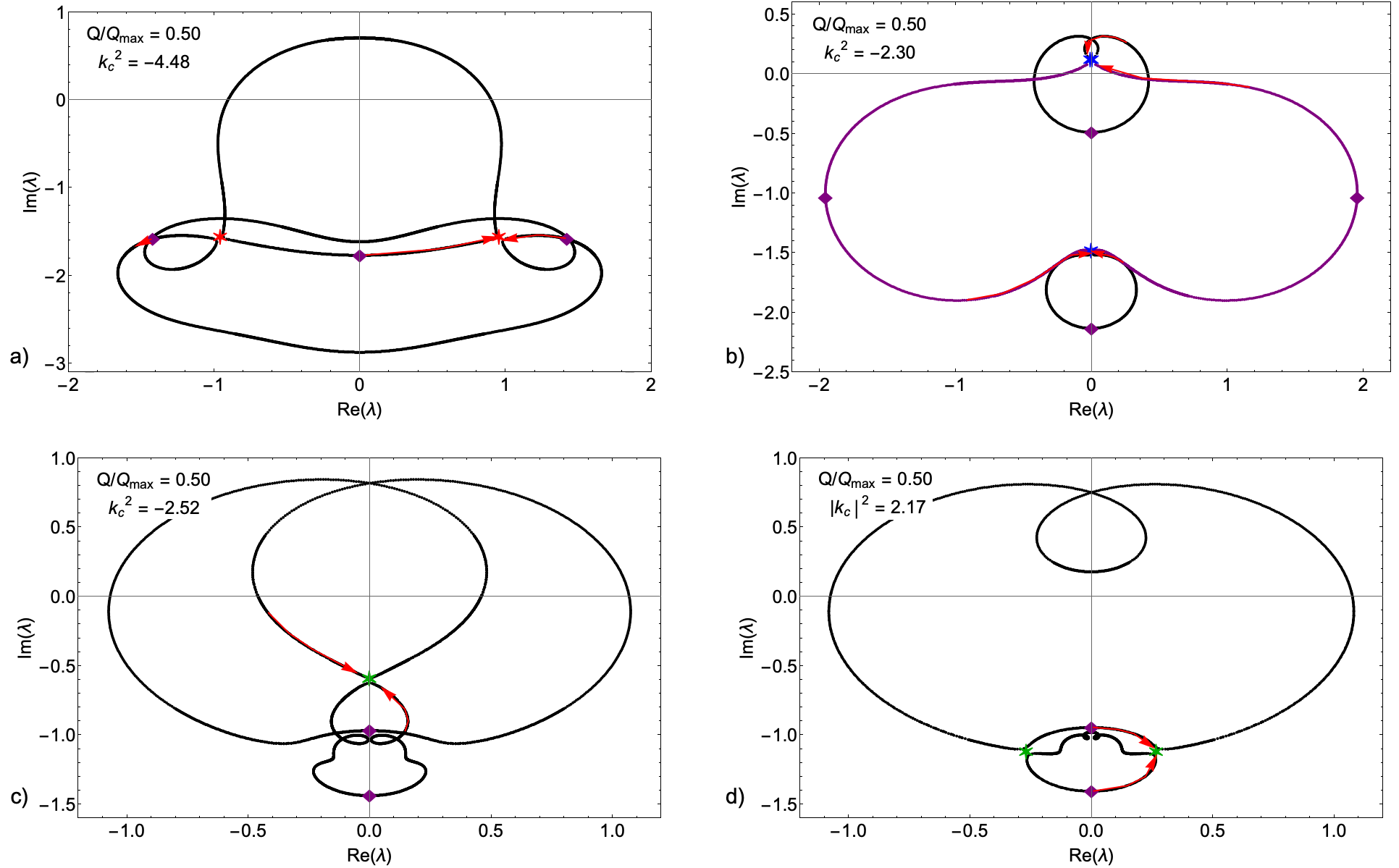}
\caption{
Quasinormal modes in the shear channel in $d=3$ at fixed $|k|^2$, illustrating the qualitatively different types of collisions.
Purple diamonds indicate the QNMs at real momentum and stars indicate the collisions, with the color matching the color in figure \ref{fig:RC2}.
The red arrows indicate the movement of the modes as the phase approaches the critical value.
}
\label{fig:shear_12}
\end{figure}

Let us now give some more details about the various branches of collisions seen in figure \ref{fig:RC2}, starting for the shear channel. As can be seen from figure \ref{fig:RC2}(a) and (b), we get qualitatively the same picture in $d=3$ and $d=4$. Thus, for the sake of brevity, we will focus on the shear mode in $d=3$.
\begin{itemize}
\item \textbf{Red branch}:
In this branch the radius of convergence is set by a collision between the shear mode and a non-hydrodynamic mode also in the shear channel.
In figure \ref{fig:shear_12}(a) we show a representative example at charge $\tilde{Q} = 0.2$ and momentum $|k|^2=|k_c|^2=4.48$. 
The purple diamonds correspond to the location of the poles for real $k$.
The purely imaginary one is the hydrodynamic shear mode, which traces the upper black curve as the phase is rotated.
The other two are a pair of non-hydrodynamic modes that trace the lower black curve.
The collision happens at the locations marked with stars, for $k^2=k_c^2=4.48\, e^{\pm\,i\pi\, 0.067}$, with the red arrows indicating the movement towards the collision at positive phase.
The branch point in the corresponding Green's function is square-root type.

\item \textbf{Blue branch}: In figure \ref{fig:shear_12}(b) we show a representative collision on the blue branch, for $d=3, \tilde Q=1/2$  and momenta $|k|^2=|k_c|^2=2.3$. The color-coding is the same as before, with the addition of the purple line which corresponds to (gapped) transverse gauge fluctuations.
We see a qualitative change in the spectrum due to a collision between the two modes occurring precisely at the locations marked with blue stars at $k^2=k_c^2=-2.3$. The collision is between the shear mode and one of a pair of non-hydrodynamic modes in the transverse channel. The other mode in the pair collides simultaneously with a non-hydrodynamic mode in the shear channel.

This type of collision was first discussed in \cite{Withers:2018srf} and the value of the critical point can actually be determined analytically by setting $\Delta_1=0$: 
\begin{align}\label{split1}
z_c^\text{shear} &=  - \frac{ \left(d - 2 + d \tilde{Q}^2 \right)^2}{2 d (d-1) \tilde{Q}^2 (1 - \tilde{Q})^2}<0\, .
\end{align}
At these values of the momentum, the shear and transverse gauge mode satisfy the same equation of motion and thus all the modes included in the two channels collide pairwise with each other. As it was also pointed out in \cite{Withers:2018srf}, this branch point is of square-root type. 

\item \textbf{Green branch}: 
In figure \ref{fig:shear_12}(c), we show the pole collision for the shear mode for $d=3$, $\tilde Q=0.9$. In this case the non-analyticity is located at $|k_c|^2=2.52$ ($k_c^2=-2.52$).
This is due to a collision between the hydrodynamic shear mode and a non-hydrodynamic mode in the shear channel. Note that the latter is purely diffusive and is different from the non-hydrodynamic mode involved in the red branch collision. At this large charge this mode has come in from minus infinity and is the dominant non-hydrodynamic mode.
This collision occurs at the location indicated by the green star and the corresponding branch cut is also of square-root type. As can be seen in figure \ref{fig:RC2}, there is a qualitative change along this branch. Even though the modes involved in this collision are the same, initially the collision takes place at imaginary $k$, while later on it switches to a complex value. This can be clearly seen in figure \ref{fig:shear_12}(d) for $d=3$, $\tilde Q=0.92$ and $|k_c|^2=2.17$ ($k_c^2=2.17 \,e^{\pm\,i\pi\, 0.118}$). 
As one approaches extremality, $|k_c|\to 0$ indicating the appearance of a branch cut at the origin along the
negative z-axis at extremality.
\end{itemize}

Let us now move on to very briefly discuss the sound channel. Just like in the shear, the radius of convergence of the hydrodynamic sound mode, shown in \ref{fig:RC2}(c) and (d), exhibits 3 distinct branches of collisions that become dominant at different values of the charge. 
\begin{itemize}
\item \textbf{Red branch:} The red branch, just like in shear, is defined as being dominated by a pole collision of the hydrodynamic sound with a gapped mode in the same channel. This collision occurs at complex values of the momentum $z$.
\item \textbf{Blue branch:} The pole collisions along the blue, analytic branch in the sound channel occur precisely at the location where $\Delta_0=0$, namely at:
\begin{align}\label{split2}
z_\text{sound} &= - \frac{ \left(d - 2 + d \tilde{Q}^2 \right)^2}{4 d (d-2) \tilde{Q}^2 (1 - \tilde{Q})^2}<0\, .
\end{align}
The defining characteristic of this branch is a collision between the hydrodynamic sound and diffusion mode, the latter of which is already beyond its radius of convergence.  As such, unlike \cite{Abbasi:2020ykq}, we wouldn't call this a hydro-hydro collision as one of the modes is well outside the validity of hydrodynamics.
\item \textbf{Green branch:} This branch is characterised by the collision of the hydrodynamic sound mode with a gapped, diffusive mode in the same channel. This collision happens at $z<0$ and corresponds to a two-sheet branch point.
\end{itemize}

\begin{figure}[h!]
\centering
\includegraphics[width=0.65\linewidth]{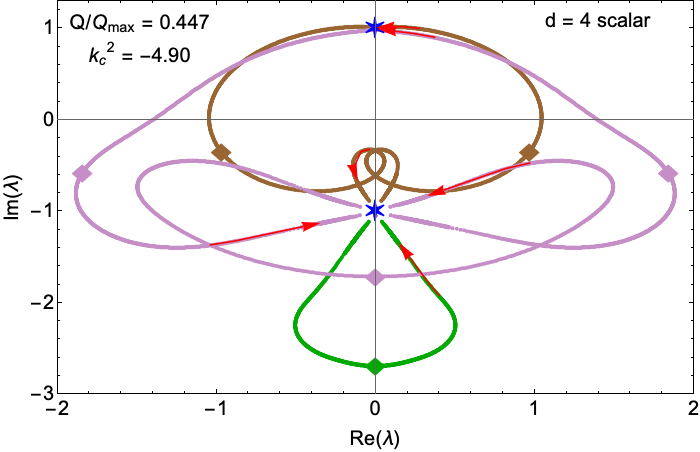}
\caption{Spectral curve for the sound hydrodynamic mode for $d=4, \tilde Q=Q/Q_{max}=0.445$, at fixed $|k|^2 = |k_c|^2 = 4.90$, which corresponds to the transition point between the blue branch and the green branch. 
The brown line corresponds to the hydrodynamic sound mode, the green line to a diffusive non-hydrodynamic mode in the sound channel and the lilac line to the diffusion mode. 
The pole collisions are indicated by the blue stars. 
At the critical point located at $\lambda=-i$ we see a collision between 4 QNMs implying a multiplicity 4 branch point.}
\label{fig:blueGreen}
\end{figure}

For the sound channel, rather than discussing every type of collision in detail, we highlight a particularly interesting one.
In $d=4$ dimensions, this happens at $\tilde Q=0.445$, which is plotted in great detail in figure \ref{fig:blueGreen}. 
The brown line corresponds to the hydrodynamic sound mode, the green line to a diffusive non-hydrodynamic mode in the sound channel and the lilac line to the diffusion mode. 
As the red arrows indicate, at $\lambda=-i$, four QNMs collide simultaneously at the lower blue star. As such, we expect that point to correspond to a 4-sheet critical point, unlike all other non analyticities shown so far. 
This happens at every transition between two branches of collisions.
Another pole collision occurs at $\lambda=+i$.

Last but not least we also have the charge diffusion channel. Just like for shear and sound, the red branch corresponds to a collision of the hydrodynamic diffusion with a gapped mode in the same channel and the blue branch corresponds to the analytic branch \eqref{split2} where the mode of interest collides with sound, which is already beyond its convergence radius, and the green branch\footnote{This branch was originally missed. We thank Daniel Area, Richard Davison, Blaise Gouteraux and Kenta Suzuki for pointing this out. \cite{Toappear}} corresponds to the collision of the hydrodynamic charge diffusion mode with a gapped, diffusive mode in the same channel.

 Before moving on, let us emphasise that there are several other pole collisions that take place at lower values of momenta that involve only non-hydrodynamic modes. These do not play any role when determining the radius of convergence of the collective excitations in the system, but they contain information about the convergence of the small-$k$ expansion of non-hydrodynamic mode dispersion relations and are important if one wants to resum the non-hydrodynamic modes.
\section{Perturbative calculation of radius of convergence}\label{sec:Pert}
Given the complicated structure of the complex QNMs in the presence of charge, we cross verify the results presented in figure \ref{fig:RC2} through a perturbative calculation. In particular, we compute the small-k expansion of the dispersion relation for the hydrodynamic shear and sound modes as in \eqref{eq:dispersions}, to high order by solving iteratively the equations of motion for the corresponding fluctuations order by order. With the perturbative coefficient of the dispersion relations at hand,  we can then extract the radius of convergence and compare with the results of the previous section. We have checked both $d=3$ and $d=4$ and we find excellent quantitative agreement, as shown already in figure \ref{fig:RC2}. This section is an extension of the calculation  presented in \cite{Withers:2018srf}.

\subsection{Shear}
For the shear mode we perform an expansion in even powers of the momentum $k$ for both the master field and the frequency 
\begin{align}
&\lambda= \sum_{n=1}^{N} \lambda_n k^{2n}\,,\nonumber\\
&\Psi_{1-}=\sum_{n=0}^{ N} \psi_n k^{2n}\,,
\end{align}
up to a large but finite order $N$. 

In the IR we again impose regularity:
\begin{equation}
\psi_n= \alpha^{(0)}_n+\alpha^{(1)}_n(1-u)+ \dots\, ,
\end{equation}
where using linearity of the equation we set:
\begin{equation}\label{eq:bchorshear}
\alpha^{(0)}_n=
\begin{cases}
1\,,&\,n=0\,\\
0,& \,n=1\,.
\end{cases}
\end{equation}

In the UV,  after rescaling $\Psi_{1-}$ by a power of $u$ to set the leading power to zero, we have the expansion:
\begin{equation}
\psi_n= \psi^{J}_n + \psi^{V}_n u^{d-2}+\dots \,,
\end{equation}
where the first term corresponds to the source and the second to the vacuum expectation value.
We want to find a solution with vanishing source, which at a given order only exists for the appropriate $\lambda_n$. 
The first order can easily be solved as $\psi_1(u) = u^{d-2}$. In order to solve the higher orders, we provide a guess  for $\lambda_n$ and then solve the equation subject to boundary condition (\ref{eq:bchorshear}).
The equation then fixes the source, which we iteratively set to zero using a root finding algorithm by varying the coefficient $\lambda_n$.

 
 With the coefficients $\lambda_n$ at hand, one can extract the radius of convergence from the behaviour of $|\lambda_n|$ at high values of $n$. 
 In particular, the slope of this function in a logarithmic scale contains information about $|k_c|^2$, while its periodicity reflects the argument. 
 In figure \ref{fig:pert}(a) we plot theses coefficients for the shear mode for $\tilde Q=0.5$, $d=4$, along with the fit we used for extracting $k_c^2$ indicated by the orange line. 
 As one can see from the figure our numerical approach is powerful enough to extract up to $N=720$ perturbative coefficients, allowing us to calculate the radius of convergence very accurately. 
 To be specific, we find the radius of convergence extracted perturbatively to be $|k_c|^2=6.01$, as opposed to $|k_c|^2=6$ obtained analytically. This corresponds to a percentage error of $0.2\%$. 
 
 Repeating this process, we scan the whole range of $\tilde Q$ for both $d=3,4$ and we already included our results in figure \ref{fig:RC2}(a)-(b) with the circles \textemdash for these data we only went up to order $N=120$. We see a very good quantitative agreement with the locations of the lowest lying branch points extracted from the spectral curves. Theoretically, given the existence of a branch point on the complex imaginary plane, the Puiseux Theorem (under the condition that $F(\omega,q)$ is analytic at $\omega=0$) implies that the radius of convergence should be set by the locations of that point. However, 
there is no formal proof that, in general, the radius of convergence should be set by a branch point.

\subsection{Sound}
We now repeat the perturbative analysis for the sound mode. In this case we found it more convenient to work with the gauge invariant fields of \cite{Kovtun:2005ev}, which we call $\Phi_H,\Phi_A$. Once again, we do a small-k expansion for the fields and the frequency of the form
\begin{align}
&\lambda= \sum_{n=1}^{N} \lambda_n k^{n}\,,\nonumber\\
&\Phi_H=\sum_{n=0}^{N} h_n k^{n}\,,\quad \Phi_A=\sum_{n=0}^{N} a_n k^{n}\,.
\end{align}
Plugging this into the equations of motion we obtain a set of $N$ nested pairs of second order linear equations, that we solve numerically. 

The method is very similar to the shear case.
In the IR, we again impose regularity
\begin{align}
h_n= h^{(0)}_n+h^{(1)}_n (1-u)+\dots\,\nonumber\\
a_n= a^{(0)}_n+a^{(1)}_n (1-u)+\dots
\end{align}
where we use the linearity of the equations to set
\begin{equation}
h^{(0)}_n=
\begin{cases}
1\,,&\,n=0\,\\
0,& \,n=1\,.
\end{cases}
\end{equation}

In the UV, again after an appropriate rescaling by a power of $u$, the asymptotic solution is:
\begin{align}
h_n&= e_n^J + \dots + e^{V}_n u^{d} +\dots ,\nonumber\\
a_n&= f_n^J + \dots + f^{V}_n u^{d-2} + \dots \,,
\end{align}
where $e_n^J , f_n^J$ correspond to the sources $e^{V}_n, f^{V}_n$ to the vacuum expectation values.

Compared to the shear we have one more equation, and correspondingly one more parameter in the IR which is $a^{(0)}_n$.
In order to solve this boundary condition problem we use the same methodology  as in the perturbative calculation for the shear, where we now tune $\lambda_n$ and $a^{(0)}_n$ to set the two sources to zero.

\begin{figure}[h!]
\centering
\includegraphics[width=1\linewidth]{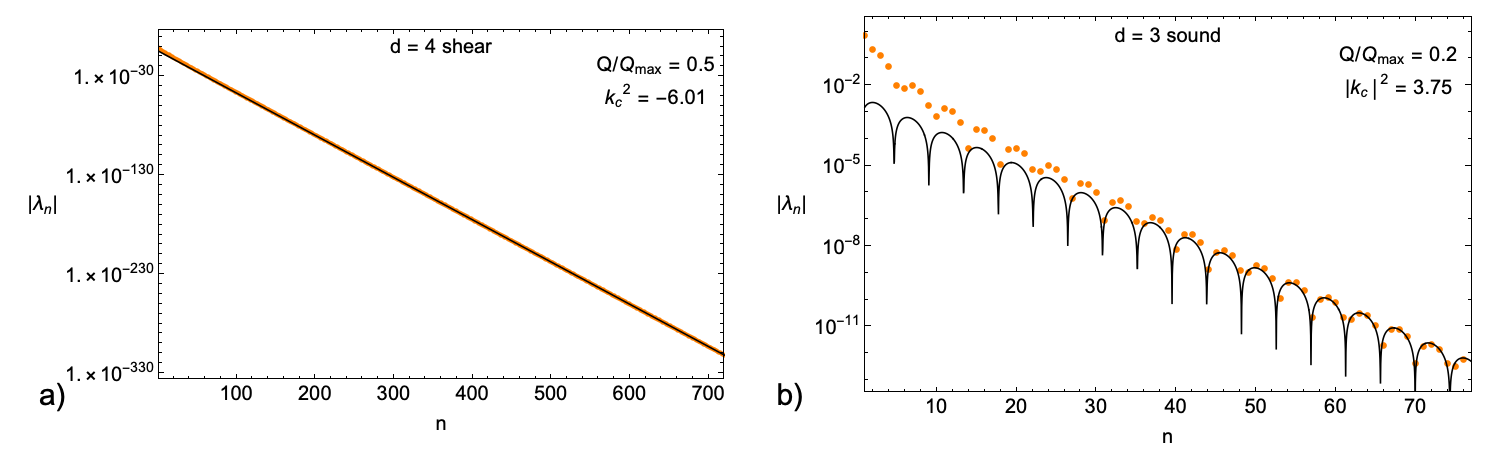}
\caption{Perturbative coefficients of the small-$k$ expansion of the dispersion relation of the shear mode for $\tilde Q=Q/Q_{max}=0.5, d=4$ (panel a) and the sound mode for $\tilde Q=Q/Q_{max}=0.2, d=3$ (panel b). By fitting appropriately these coefficients, as depicted by the solid black lines, we are able to extract the corresponding radius of convergence to be $k_c^2=-6.01$ in panel (a) and $k_c^2= 3.75\, e^{\pm\,i\pi\, 0.054}$ in panel (b).}
\label{fig:pert}
\end{figure}

In figure \ref{fig:pert}(b) we plot the resulting coefficients $|\lambda_n |$  for the sound mode for $\tilde Q=0.2$, $d=3$ on a logarithmic scale, where the orange line indicates the fit we used for extracting $k_c^2$. The oscillations observed are indicative of the collision taking place at complex momenta. 

Our results for other values of $\tilde Q$ and $d$ can already be seen in figure \ref{fig:RC2}(c)-(d). Note that in the case of sound, the numerical calculation was much come computationally demanding compared to the case of shear. This is reflected in the fact that we weren't able to go to high enough $N$ within precision and thus the fits weren't as accurate. Nevertheless, we see good agreement with the results of the previous section.

\section{Pole skipping for charged black branes}\label{sec:poleSkip}

In this section we investigate the phenomenon of pole skipping in the presence of charge. This phenomenon occurs at points on the imaginary frequency and momentum plane where lines of zeros and poles of the corresponding retarded Green's function overlap \textemdash these are to be contrasted with the points analyzed in section \ref{sec:NonP}  which correspond to overlaps of two lines of poles, one of which corresponds to the hydrodynamic mode. As such, the various correlators are not uniquely defined at these points but instead they are infinitely multivalued, depending on how the point is approached.

As it was pointed out in \cite{Blake:2018leo,Blake:2019otz}, the value of the frequency and momentum where pole skipping occurs can be extracted directly by studying the IR behaviour of the gravitational system and identifying special point in the $(\lambda,z)$ plane where the IR expansion is underdetermined and thus there is no unique solution that is ingoing at the horizon. Following this approach one can identify two types of pole skipping points, depending on whether they appear in the upper or lower frequency plane. In particular, in the upper half frequency plane there is only one such point, which is located in the sound channel. As such it is associated with the exponential growth of the energy-energy correlator and thus chaos. On the other hand, in the lower frequency plane we have pole skipping points at the Matsubara frequencies in all channels. These are not expected to be related to chaos \cite{Blake:2019otz}.

Other than its connection with chaos, this phenomenon is particularly interesting as it provides an infinite set of analytic constraints on the dispersion relations of the modes of the system and consequently the correlators themselves.

\subsection{Matsubara frequencies}
In this subsection, we show that at the Matsubara frequencies, $\lambda_m=-i \,m$ with $m\neq0$ (or $\omega_m = - i m 2 \pi T$), and at specific values of the momentum $z^{(m)}_*$, all channels (including the ones that do not contain any hydrodynamic modes) exhibit the phenomenon of pole-skipping. The values of $z^{(m)}_*$ are determined from the determinant of an $m \times m$ matrix $M^{(m)}$ obtained from the near-horizon expansion of the master fields equations,  as prescribed in \cite{Blake:2019otz}.

In more detail, in Eddington-Finkelstein coordinates, imposing ingoing boundary conditions at the horizon boils down to requiring regularity of the fluctuations at $u=1$. We thus assume that each of the master fields admits an IR expansion of the form
\begin{align}
&\Phi=\sum_{m=0}^{\infty} c_m(1-u)^m\,.
\end{align}
Plugging this expression in the master equation and expanding close to the horizon, the equation take the following form
\begin{equation}
M(\lambda,z)\cdot c\equiv
\begin{pmatrix}
M_{11} &(\lambda+i)& 0&0& \dots \\
M_{21} &M_{22}& (\lambda+2i)&0& \dots \\
M_{31} &M_{32}&  M_{33}&(\lambda+3i)& \dots \\
 \dots  &  \dots &  \dots & \dots \\
\end{pmatrix}
\begin{pmatrix}
c_0 \\
c_1 \\
c_2 \\
 \dots\\\,
\end{pmatrix}
=0
\end{equation}
where $M_{ij}=M_{ij}(z,\lambda,\tilde Q)$ \textemdash the specific formulas are lengthy and not very illuminating and thus have been omitted. For frequencies $\lambda\neq \lambda_m$ it is easy to iteratively solve this system of algebraic equations, fixing the coefficients $c_1,c_2,\dots$ in terms of $c_0$. However, for $\lambda=\lambda_m$ this is not always possible: there are certain values of the momenta where not only $c_0$ but also the coefficient $c_m$ remains undetermined, leading to multiple ingoing solutions. To identify these values of the momentum we follow the criterion of \cite{Blake:2019otz}
\begin{align}
\label{eq:pscond}
det M^{(m)}=0\,,\nonumber\\
\sqrt{z}\, \partial_z det M^{(m)}\ne0\,,
\end{align}
where we define $M^{(m)}$ as the matrix constructed by the first $m$ rows and $m$ columns of $M$. The first criterion corresponds to the matrix $M^{(m)}$ being non-invertible leading an underdetermined system for the coefficients of the near-horizon expansion, while the second one excludes anomalous points. In this context an anomalous point is a point where, even though the near-horizon expansion is not uniquely fixed, the corresponding Green's function does not take the appropriate pole skipping form characterized by the overlap of a series of poles and a series of zeros. Note that the second constraint essentially demands that the pole skipping points are of multiplicity one. Following the methodology above, we now present our results for the three sectors of interest:\\

\noindent\textbf{\underline{Tensor sector ($d=4$):}} In this sector, we have identified $m$ pole skipping points at each Matsubara frequency 
 \begin{equation}
 \lambda_m=-i\,m\,,\quad z^{tensor (m)}_*=\{z^{(m)}_1, \dots, z^{(m)}_m\}\,.
 \end{equation}
The expressions for $z_*$ are very complicated and are thus omitted. Note that at every level $m$, there is one value of $z_*$ that is positive.\\
 
\noindent\textbf{\underline{Vector sector:}} Just like in the tensor sector, here we also obtain pole skipping points at the Matsubara frequencies
  \begin{equation}
 \lambda_m=-i\,m\,,\quad z^{vec (m)}_*=\{z^{(m)}_1, \dots, z^{(m)}_{2m}\}\,.
 \end{equation}
However, in this case we find $2m$ such points at each level, out of which at least one is positive. Note that when $\tilde Q=0$, these decouple to $m$ solutions corresponding to the transverse metric perturbations studied in appendix E of \cite{Blake:2019otz} and $m$ to transverse gauge field perturbations\footnote{For $d=3$ and $m=1$ this solution is in fact an anomalous point and should be discarded.}.  Note however that this splitting is not even for all values of the charge: given that the pole skipping point should lie on dispersions relation and these change as functions of the charge, it is natural for the splitting across the two channels in this sector not to be always even and, in fact, to change with the charge. 
This sector was briefly discussed in appendix E.5 of \cite{Blake:2019otz} for $m=1$ in $d=4$ dimensions, and our results are in quantitative agreement.\\

\noindent\textbf{\underline{Scalar sector:}} Matsubara frequencies also appear in the sound channel
  \begin{equation}
  \begin{cases}
 \lambda_1=-i\,,\quad z^{scal (1)}_*=\{z^{(1)}_1, z^{(1)}_{2},z^{(1)}_{3}\}\,,\quad m=1\\
  \lambda_m=-i\,m\,,\quad z^{scal (m)}_*=\{z^{(m)}_1, \dots, z^{(m)}_{2m}\}\,,\quad m\ge2\,.
 \end{cases}
 \end{equation}
In the case with no charge, at $m=1$, these points split into 2 modes in the sound channel and 1 mode in the diffusion channel and match the results of \cite{Blake:2019otz}. For bigger $m$, the modes split evenly in the chargeless limit. At finite charge, the split depend on the value of $\tilde Q$. The extra solution at $m=1$ was also seen in the chargeless case of \cite{Blake:2019otz} and it's simply a consequence of the more complicated structure of the equations for the longitudinal metric perturbations.

\subsection{Pole skipping associated with chaos}
In addition to the pole skipping points discussed above, in the sound channel there is one additional point in the upper half frequency plane. It is located  at $\lambda_*=+i$ and 
\begin{equation}
z^{chaos}_*=
\begin{cases}
4/[3(\tilde Q^2-1)]\,,&\,d=3\,\\
3/[2(\tilde Q^2-1)],& \,d=4\,.
\end{cases}
\end{equation}
These values of frequency and momentum can be associated with the holographic Lyapunov exponent and the butterfly velocity, determined independently by considering shock waves in $AdS_{d+1}$.These expressions agrees with the results of \cite{Grozdanov:2019uhi} in the case with no charge, $\tilde Q=0$, in $d=4$ and are associated with an exponential growth of the retarded two-point function related to chaos.

In comparing $|z^{chaos}_*|$  with the positive element in $z_*^{vec(1)}$ (which always lie on the hydrodynamic shear dispersion), we see that the two are equal to each other only when there is no charge in the system (for both $d=3,4$). This supports the conclusion of \cite{Blake:2019otz} that the pole-skipping points in the lower half plane are in general not fixed by the butterfly velocity and are thus not related to chaos in a fundamental way.

It is interesting to note that in the extremal limit, the value of the momentum associated to chaos goes to zero
\begin{equation}
\lim_{\tilde Q\to1} \frac{(2\pi T)^2}{\mu^2}|z_*^{chaos}|=0\,.
\end{equation}
However, this result should be taken with a grain of salt as it is not clear a priori that the near-horizon and extremal limit commute and thus a direct IR analysis in the extremal limit might yield a different result. It would be interesting to confirm this with a direct calculation at extremality.

\subsection{Comments}
There are three points that we would like to emphasis:
\begin{itemize}
\item It is astonishing how much non-trivial information about the Green's functions is stored in the pole skipping points: they provide an infinite set of constraints on the dispersion relations of the poles of the corresponding correlators. To this end, we have  checked the relationship between the dispersion relations of the five modes in the system and the pole skipping values for the chaos point and for the Matsubara frequencies up to $m=5$. For example in figure \ref{fig:ps} we show such a comparison for the vector and scalar channels for $z$ real. In particular in panel (a), focusing on $d=3,\tilde Q=0.9$, we plot the dispersion relations of the shear channel with orange and the dispersion for the transverse gauge with black. The corresponding Matsubara points are indicate with the red and purple points respectively and we see a perfect match. Similarly, in panel (b) the orange lines correspond to the dispersion of the charge diffusion and the black lines to the sound channel $d=4,\tilde Q=0.9$. Once again we see a perfect agreement with the red and purple points respectively. Focusing on the hydrodynamic sound mode dispersion, we see that other than the chaos point that is located in the upper half frequency plane (and not explicitly shown in this figure), no other Matsubara point lies along its dispersion. 
On the contrary, the hydrodynamic shear and diffusion dispersions have a pole skipping point on them at \textit{every} Matsubara frequency.

\begin{figure}[h!]
\centering
\includegraphics[width=1\linewidth]{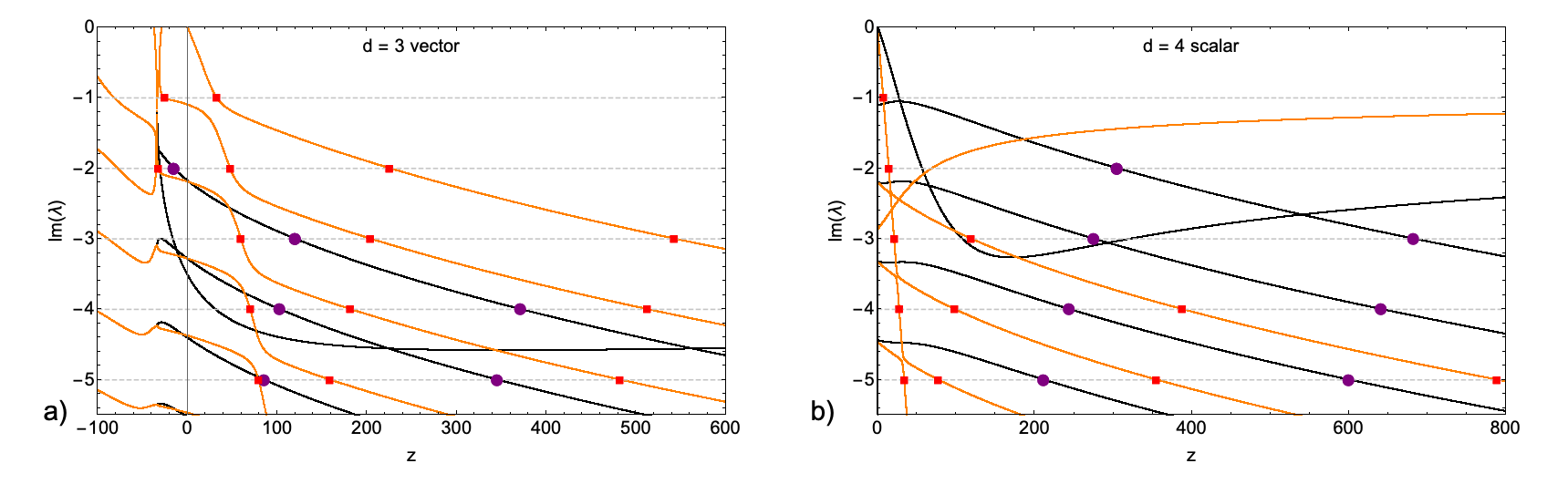}
\caption{Comparison of the pole skipping points at the Matsubara frequencies with the full dispersion relations for real $z$. 
Panel (a): The orange (black) lines correspond to the shear (transverse) channel for $d=3, \tilde Q=Q/Q_{max}=0.9$. 
Panel (b): The orange (black) lines correspond to the diffusion (sound) channel for $d=4, \tilde Q=Q/Q_{max}=0.9$. 
In both panels, the red and purple points correspond to the locations of pole skipping and it is easy to see they all lie along dispersion relations, either of hydrodynamic or non-hydrodynamic QNMs. Note that complex pole skipping points are not visible on this plots.}
\label{fig:ps}
\end{figure}

\item In figure \ref{fig:pstest} (a) we compare the radius of convergence for the sound mode for $d=4$ with  $|z_*^{chaos}|$, which is depicted by the brown line. We see that for $\tilde Q>0.45$, the radius of convergence is smaller than the pole skipping value that is associated to chaos, while the opposite is true for $\tilde Q<0.45$. 
We conclude that the two are unrelated, which is in agreement with the conclusion of \cite{Grozdanov:2019uhi}. We also carry out a similar comparison for the shear channel. In particular, in figure \ref{fig:pstest} (b) we indicate the two solutions  in $|z_*^{vec(1)}|$ in brown, where the segments plotted with a solid lines correspond to pole skipping points that lie along the shear dispersion \footnote{Note that the solid brown line at higher values correspond to the hydrodynamic shear mode, while the other one to a non-hydrodynamic mode in the same channel}. We conclude that there is no strict relationship between pole skipping and radius of convergence in this channel either.

\item In the sound channel, of particular interest  are the cases $(\tilde Q,|k|^2)=(0.5,4.5)$ for $d=4$ and $(\tilde Q,|k|^2)=(1/\sqrt{5},2.4)$ for $d=3$, at which the analytic branch intersects the chaos point. The first of these cases is plotted in figure \ref{fig:analPS}. 
The brown line corresponds to the hydrodynamic sound mode, the green line to a diffusive non-hydrodynamic mode in the sound channel and the lilac line to the diffusion mode. 
At this value of $|k|^2$ we are slightly above the radius of convergence. The sound mode has already collided with the diffusive non-hydrodynamic mode at $|k|^2=|k_c|^2=1.95$ and subsequently, at $|k|^2=2$, it undergoes a second collision. In particular, in figure \ref{fig:analPS}  we see three pair-wise pole collision (compare to figure \ref{fig:blueGreen}). 
One of these occurs precisely at the chaos pole skipping point at $\lambda=+i$ and purely imaginary $k$ \textemdash we indicate this point with an orange circle.

\begin{figure}[h!]
\centering
\includegraphics[width=1\linewidth]{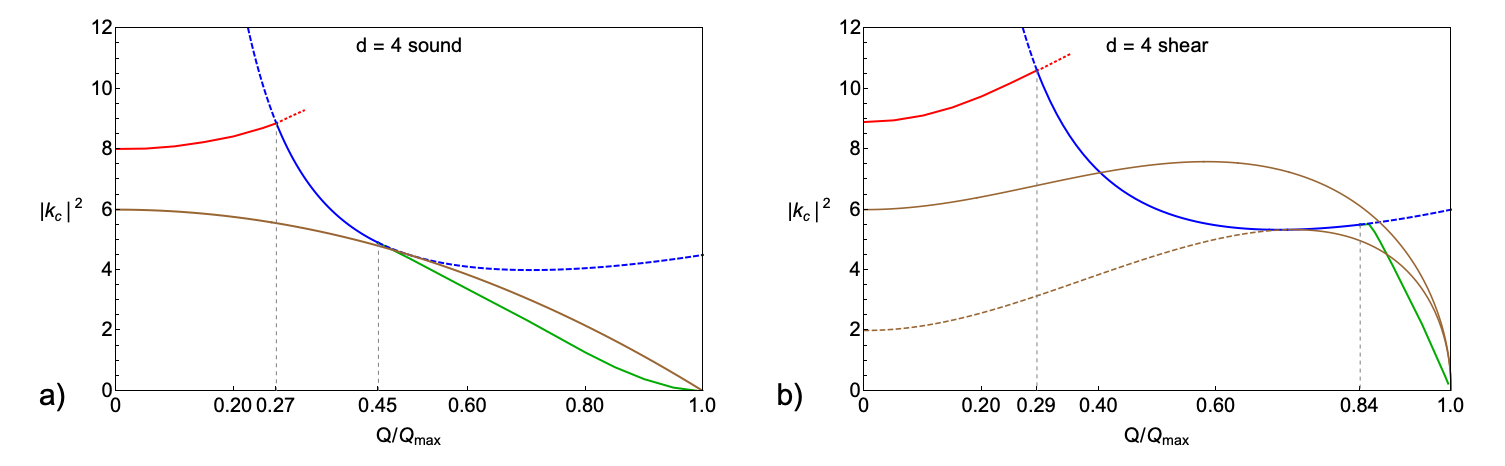}
\caption{Panel (a): Comparison of the radius of convergence for the sound mode for $d=4$ with $|z_*^{chaos}|$ as a function of $\tilde Q=Q/Q_{max}$. 
Panel (b): Comparison of the radius of convergence for the shear mode for $d=4$ with the two solutions in $|z_*^{vec(1)}|$, which are depicted in brown. 
The top brown line corresponds to pole skipping along the hydrodynamic shear mode dispersion. The bottom brown curve correspond to pole skipping points along the transverse gauge mode (dashed segment) and a gapped mode in the shear channel (solid segment). In both panel (a) and (b) we see that as the charge in the system is varied, the relationship between the two quantities changes. This leads us to conclude that the two are unrelated.}
\label{fig:pstest}
\end{figure}

\begin{figure}[h!]
\centering
\includegraphics[width=0.65\linewidth]{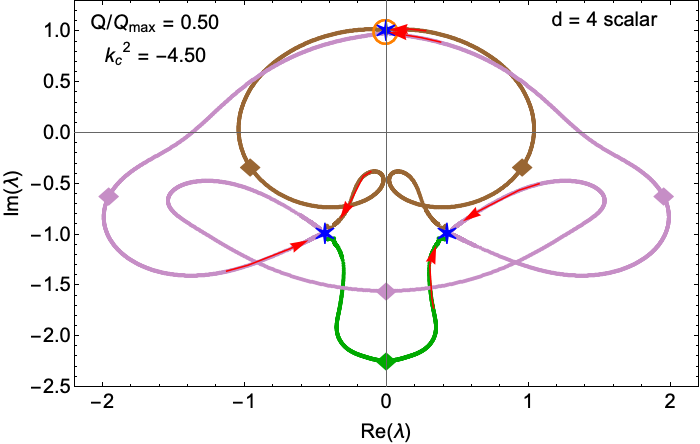}
\caption{Spectral curve of the hydrodynamic sound mode on the complex frequency plane for $d=4,\tilde Q=Q/Q_{max}=0.5$ at fixed $|k|^2=4.5$. 
The brown line corresponds to the hydrodynamic sound mode, the green line to a diffusive non-hydrodynamic mode in the sound channel and the lilac line to  the diffusion mode. 
This value of the momenta is beyond the radius of convergence and, in fact, lies along the subdominant blue branch. 
This is precisely the reason why we see collisions corresponding to the blue \textit{and} the green branch at the locations indicated by the stars. 
The pole skipping point in marked with an orange circle and clearly lies along the hydrodynamic sound dispersion. }
\label{fig:analPS}
\end{figure}

\end{itemize}
\section{Discussion}\label{sec:concl}
In this work we have extracted the radius of convergence of the collective modes of a charged fluid in a microscopic theory dual to the RN black brane in $d=3,4$ dimensions using two different methods. The first one involves the study of spectral curves on the complex frequency  and momentum plane, while the second one involves a perturbative expansion in small momenta up to a very large order. The two methods give results that are in good quantitative agreement. Our results clear up the apparent confusion in the literature regarding the analytic branch of \cite{Withers:2018srf} and its compatibility with the results of \cite{Grozdanov:2019uhi,Grozdanov:2019kge} for the neutral case. As shown is section \ref{sec:NonP}, this type of pole collision becomes subdominant in the limit of no charge and as such, does not set the radius of convergence. Instead, the latter is set by a different type of collision that occurs at a finite value of the momenta, rendering the radius of convergence finite at $\tilde Q=0$. 
In the opposite limit, $\tilde Q \rightarrow1$, corresponding to the extremal case, the radius of convergence of all hydrodynamic modes, which is set by different type of collision, vanishes in the limit. For the diffusion mode it remains finite. These results match nicely the findings of \cite{Edalati:2010hk,Edalati:2010pn} which indicate that corelators associated to the shear and sound collective excitations exhibit branch cuts along the negative imaginary frequency axis.

We also carried out a detailed analysis of the phenomenon of pole skipping in the case of RN black branes in both $d=3$ and $d=4$ dimensions. Extending the results of \cite{Blake:2019otz}, we obtained a set of pole skipping points on the lower frequency plane corresponding to Matsubara frequencies. We also identified one such point on the upper half frequency plane, related to the exponential growth of the energy-energy two-point function and associated to quantum chaos, extending the results of \cite{Grozdanov:2017ajz}  to finite charge. We have tested and confirmed that all these points lie along dispersion relations of hydrodynamic or non-hydrodynamic modes, providing an infinite set of constraints for the dispersion relations of these modes. Finally, we compared the radius of convergence in the sound and shear channels with the pole skipping points associated with chaos and with the lowest Matsubara frequency in the vector channel, as functions of the charge. We concluded  that  there is no fundamental connection between the two: depending of the value of $\tilde Q$, the pole-skipping point can be within or outside the radius of convergence of hydrodynamics. 

As we explained in the main text, our results for the extremal limit for pole skipping should be taken with caution. This is because it is not clear a priori if the extremal limit and the near-horizon expansion commute. It would be interesting to repeat this computation starting directly at zero temperature. Furthermore, it is interesting to establish how generic are our results for the radius of convergence and whether or not they apply to other holographic systems with finite chemical potential such as the STU model \cite{Son:2006em}. One could also attempt to extend the formalism of \cite{Heller:2020uuy} to the extremal case as well as to the case with non-zero charge.

Finally, it would be interesting to exploit our perturbative results in order to investigate further the possibility of Pade-resummation in the sound channel in order to go beyond the branch point singularity and recover some of the non-hydrodynamic modes, along the lines of \cite{Withers:2018srf} .

\section*{Acknowledgements}
We would like to thank Petar Tadic and especially Benjamin Withers for helpful discussions.  The work of AJ is supported by ERC Advanced Grant GravBHs-692951.  C.P. is supported by the European Union’s Horizon 2020 research and innovation programme under the Marie Skłodowska-Curie grant agreement HoloLif No 838644. 
\appendix

\section{Master Equations}\label{app:MasterEquations}
In this appendix we derive the UV boundary conditions (corresponding to the QNM condition of vanishing sources at the boundary) for the master scalar in the scalar channel in  $d=3$. To do this we need to know how the master scalar is related to the gauge invariant combinations.
To keep the discussion brief we give only the essential details here and refer the interested reader to~\cite{Jansen:2019wag} for more information.
The two gauge invariants we focus on are $H_{tt}$ and $A_t$, which are built respectively on the fluctuations $\delta g_{tt}$ and $\delta a_t$ with corrections from other components to make them invariant under infinitesimal gauge and coordinate transformations.
There are four more gauge invariant combinations, which can be related to these two using the perturbation equations.

Near the boundary these two invariants fall off as
\begin{align}\label{eq:ginvasymps}
H_{tt} &= J_H u^{-2} - \frac{3 i \lambda}{2} (1 - \tilde{Q}^2) J_H u^{-1} - \frac{9 z}{8} (1 - \tilde{Q}^2)^2 J_H  +  O_H u  \\
&- \Big[ \frac{3 i \lambda}{8} (1 - \tilde{Q}^2) \left( (1+3\tilde{Q}^2) J_H + 4 \sqrt{3} \tilde{Q} J_A + 4 O_H \right) \\
&+ \sqrt{3} \tilde{Q} O_A + \frac{81 z^2}{128} (1 - \tilde{Q}^2)^4 J_H  \Big]u^2 + \mathcal{O}(u^3)\, , \\
A_t &= J_A + O_A u - \frac{3}{8} (1 - \tilde{Q}^2) \left( 3 J_A z (1 - \tilde{Q}^2) + 4 i \lambda O_A \right) u^2 + \mathcal{O}(u^3)\, ,
\end{align}
where $J$ and $O$ are the sources and vacuum expectation values.
The QNM boundary conditions are $J_H = J_A = 0$. 

To see what this means for the master scalars, we invert the map giving the gauge invariants in terms of the master equations of~\cite{Jansen:2019wag}.
Plugging in the asymptotics of Eq. (\ref{eq:ginvasymps}) with the sources set to zero into that we obtain
\begin{align}
\Phi_H &=- \left(\frac{8 O_H}{27 (1 - \tilde{Q}^2)^4 (z - \lambda^2)^2} \right) u \\
& \left(\frac{4(8+24 \tilde{Q}^2 - 9 i \lambda z (1 - \tilde{Q}^2)^3) O_H + 32 \sqrt{3} (z-\lambda^2) \tilde{Q} (1 - \tilde{Q}^2)^2 O_A}{81 (1 - \tilde{Q}^2)^6 z (z - \lambda^2)^2 } \right) u^2 + \mathcal{O}(u^3)\, , \\
\Phi_A &=  -\left(\frac{8 O_A}{27 (1 - \tilde{Q}^2)^3 \sqrt{z} (z - \lambda^2)}  \right) u \\
&- \left(  \frac{ 8 \sqrt{3} \tilde{Q} O_H + 9 i \lambda (z - \lambda^2) (1 - \tilde{Q}^2)^3 O_A}{81 (1 - \tilde{Q}^2)^5 \sqrt{z} (z - \lambda^2)^2 } \right) u^2 + \mathcal{O}(u^3) \, .
\end{align}
If we rescale both master scalars as $\Phi = u \Phi^f$ we can identify the first term with $\Phi^f(0)$ and the second with $\Phi^{f \prime}(0)$,  we obtain the mixed boundary conditions
\begin{align}
\Phi^{f \prime}_H(0) &= - \frac{3 i \lambda}{2} (1 - \tilde{Q}^2) \Phi^{f}_H(0) -\frac{4}{3z} \frac{1+3\tilde{Q}^2}{(1-\tilde{Q}^2)^2} \Phi^{f}_H(0)
- \frac{4}{\sqrt{3z}} \frac{\tilde{Q}}{1 - \tilde{Q}^2} \Phi^{f}_A(0) \, , \\
\Phi^{f \prime}_A (0) &= - \frac{3 i \lambda}{2} (1 - \tilde{Q}^2) \Phi^{f}_A(0)  - \frac{4 \tilde{Q}}{\sqrt{3 z}(1 - \tilde{Q}^2)} \Phi^{f}_H(0) \, .
\end{align}

Finally the equations are decoupled by taking the linear combinations
\begin{equation}
\Phi_{0\pm} = \Phi_A^f + \frac{1 + 3 \tilde{Q}^2 \pm \Delta_0}{\sqrt{8 z} \tilde{Q} (1 - \tilde{Q}^2)} \Phi_H^f \, .
\end{equation}
This also decouples the boundary conditions, which become 
\begin{equation}
\Phi_{0\pm}^\prime(0) = - \frac{3 i \lambda}{2} (1 - \tilde{Q}^2) \Phi_{0\pm}(0) - \frac{2(1 + 3 \tilde{Q}^2 \pm \Delta_0)}{3 z (1 - \tilde{Q}^2)^2}\Phi_{0\pm}(0) \, .
\end{equation}

\bibliographystyle{utphys}
\bibliography{bib}{}

\end{document}